\documentclass[12pt,a4paper]{article}
\usepackage[margin=0.8in]{geometry}
\usepackage{pgfplots}
\usepackage{natbib}
\usepackage{booktabs}
\usepackage[center]{caption}
\usepackage{algpseudocode}
\usepackage{algorithm}
\usepackage{etex}
\usepackage{tikz}
\usetikzlibrary{patterns}
\usepackage[fleqn]{amsmath}
\usepackage{amssymb}
\usepackage{hyperref}
\baselinestretch
\pgfplotsset{compat=1.8}

\hypersetup{
    colorlinks=true,
    citecolor=black,
    urlcolor=blue,
    linkcolor=black,
    pdfborder={0 0 0},}

\title{High-dose-rate prostate brachytherapy inverse planning on dose-volume criteria by simulated annealing}
\date{}
\author{T. M. Deist$^1$
        \qquad B. L. Gorissen$^2$ \\
\textit{\small $^1$ Tilburg University, Department of Econometrics and Operations Research,} \\
\textit{\small 5000 LE Tilburg, Netherlands}\\
\textit{\small Corresponding author. {\tt timo.deist@maastro.nl}} \\
\textit{\small $^2$ Tilburg University, Department of Econometrics and Operations Research,} \\
\textit{\small 5000 LE Tilburg, Netherlands}}

\begin{document}
\newcommand{\unit}[1]{\ensuremath{\, \mathrm{#1}}}

\maketitle
\begin{abstract}
High-dose-rate brachytherapy is a tumor treatment method where a highly radioactive source is brought in close proximity to the tumor. In this paper we develop a simulated annealing (SA) algorithm to optimize the dwell times at preselected dwell positions to maximize tumor coverage under dose-volume constraints on the organs at risk. Compared to existing algorithms, our algorithm has advantages in terms of speed and objective value and does not require an expensive general purpose solver. Its success mainly depends on exploiting the efficiency of matrix multiplication and a careful selection of the neighboring states. In this paper we outline its details and make an in-depth comparison with existing methods using real patient data.
\end{abstract}
\begin{tikzpicture}[remember picture,overlay]
\node[anchor=south,yshift=01pt] at (current page.south) {\fbox{\parbox{\dimexpr\textwidth-\fboxsep-\fboxrule\relax}{\footnotesize This is an author-created, un-copyedited version of an article published in Physics in Medicine and Biology \href{http://dx.doi.org/10.1088/0031-9155/61/3/1155}{DOI:10.1088/0031-9155/61/3/1155}.}}};
\end{tikzpicture}
\vspace{-1cm}
\section{Introduction} \label{sec:introduction}
High-dose-rate (HDR) brachytherapy is a form of radiation therapy in which the tumor is temporarily exposed to a highly radioactive source which dwells at different positions in or around the planning target volume (PTV). For prostate tumors, the dwell positions are offered by a temporary transperineal implant of catheters which run through the prostate. A remote afterloader, which is connected to all the catheters, then sends a radioactive source through the catheters one by one, stopping at several dwell positions inside the PTV. We assume that the catheter locations and the dwell positions are known. The goal of the treatment planner is to irradiate the PTV while sparing the surrounding organs at risk (OARs) by optimizing the dwell time at each dwell position. Traditionally, treatment planning was a forward process, where dwell times were adjusted until the dose distribution was satisfactory. In the last fifteen years, inverse planning has been developed as a computerized technique where the dwell times are optimized according to the treatment planner's preferences on the dose distribution.

In the remainder of this introduction, we explain how dose distributions are evaluated, we provide a literature review on the ongoing progress of inverse planning, and we show how this work contributes to these developments.
\newpage

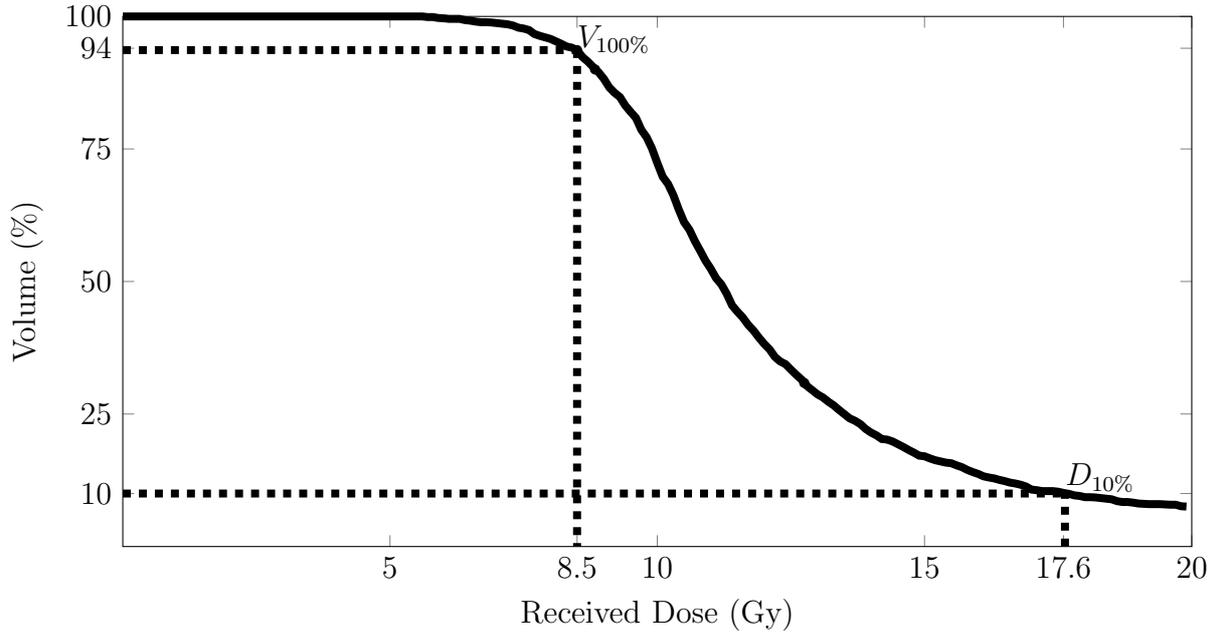
\begin{figure}
\caption{Dose-volume histogram for a prescribed dose of $8.5$ Gy.}
\label{fig:dvhexample}
\begin{center}
% This file was created by matlab2tikz v0.4.0.
% Copyright (c) 2008--2013, Nico Schlömer <nico.schloemer@gmail.com>
% All rights reserved.
% 
% The latest updates can be retrieved from
%   http://www.mathworks.com/matlabcentral/fileexchange/22022-matlab2tikz
% where you can also make suggestions and rate matlab2tikz.
% 
% 
% 
\begin{tikzpicture}

\begin{axis}[%
width=400pt,
height=200pt,
clip=false,
scale only axis,
xmin=0,
xmax=20,
xtick={    5,   8.5,    10,    15, 17.6,    20},
xlabel={Received Dose (Gy)},
ymin=0,
ymax=100,
ytick={   10,    25,    50,    75, 94,   100},
ylabel={Volume (\%)}
]
\addplot [
color=black,
solid,
line width=3.0pt,
forget plot
]
table[row sep=crcr]{
0.001 100\\
0.101 100\\
0.201 100\\
0.301 100\\
0.401 100\\
0.501 100\\
0.601 100\\
0.701 100\\
0.801 100\\
0.901 100\\
1.001 100\\
1.101 100\\
1.201 100\\
1.301 100\\
1.401 100\\
1.501 100\\
1.601 100\\
1.701 100\\
1.801 100\\
1.901 100\\
2.001 100\\
2.101 100\\
2.201 100\\
2.301 100\\
2.401 100\\
2.501 100\\
2.601 100\\
2.701 100\\
2.801 100\\
2.901 100\\
3.001 100\\
3.101 100\\
3.201 100\\
3.301 100\\
3.401 100\\
3.501 100\\
3.601 100\\
3.701 100\\
3.801 100\\
3.901 100\\
4.001 100\\
4.101 100\\
4.201 100\\
4.301 100\\
4.401 100\\
4.501 100\\
4.601 100\\
4.701 100\\
4.801 100\\
4.901 100\\
5.001 100\\
5.101 100\\
5.201 100\\
5.301 100\\
5.401 100\\
5.501 100\\
5.601 100\\
5.701 99.8857142857143\\
5.801 99.8857142857143\\
5.901 99.7142857142857\\
6.001 99.6571428571429\\
6.101 99.5428571428571\\
6.201 99.5428571428571\\
6.301 99.5428571428571\\
6.401 99.3142857142857\\
6.501 99.2\\
6.601 99.0285714285714\\
6.701 98.9142857142857\\
6.801 98.9142857142857\\
6.901 98.8571428571429\\
7.001 98.7428571428571\\
7.101 98.6285714285714\\
7.201 98.5142857142857\\
7.301 98.2857142857143\\
7.401 97.8857142857143\\
7.501 97.7142857142857\\
7.601 97.4285714285714\\
7.701 96.6857142857143\\
7.801 96.3428571428571\\
7.901 96.0571428571429\\
8.001 95.7714285714286\\
8.101 95.3142857142857\\
8.201 94.8571428571429\\
8.301 94.3428571428571\\
8.401 94.1142857142857\\
8.501 93.6571428571429\\
8.601 92.3428571428571\\
8.701 91.4857142857143\\
8.801 90.4571428571429\\
8.901 89.4857142857143\\
9.001 88.2285714285714\\
9.101 86.6285714285714\\
9.201 85.6\\
9.301 84.8\\
9.401 83.2\\
9.501 82\\
9.601 80.8571428571429\\
9.701 78.6285714285714\\
9.801 77.2\\
9.901 75.0857142857143\\
10.001 72.3428571428571\\
10.101 69.7714285714286\\
10.201 68.4\\
10.301 66.2857142857143\\
10.401 63.6\\
10.501 61.2\\
10.601 59.7714285714286\\
10.701 57.6571428571429\\
10.801 55.8285714285714\\
10.901 53.9428571428572\\
11.001 52.4\\
11.101 50.6285714285714\\
11.201 49.4285714285714\\
11.301 47.6571428571429\\
11.401 45.4857142857143\\
11.501 44.2857142857143\\
11.601 43.2\\
11.701 41.8285714285714\\
11.801 40.7428571428571\\
11.901 39.4285714285714\\
12.001 38.2285714285714\\
12.101 37.2\\
12.201 35.7714285714286\\
12.301 34.9142857142857\\
12.401 34.4\\
12.501 33.3714285714286\\
12.601 32.4\\
12.701 31.4857142857143\\
12.801 30.2857142857143\\
12.901 29.4857142857143\\
13.001 28.6857142857143\\
13.101 28.1142857142857\\
13.201 27.3142857142857\\
13.301 26.6285714285714\\
13.401 25.7714285714286\\
13.501 24.9714285714286\\
13.601 24.1714285714286\\
13.701 23.7142857142857\\
13.801 23.0857142857143\\
13.901 22.1714285714286\\
14.001 21.4857142857143\\
14.101 20.9714285714286\\
14.201 20.2857142857143\\
14.301 20.1714285714286\\
14.401 19.8285714285714\\
14.501 19.3142857142857\\
14.601 18.8\\
14.701 18.2285714285714\\
14.801 17.7142857142857\\
14.901 17.1428571428571\\
15.001 17.0285714285714\\
15.101 16.6285714285714\\
15.201 16.2857142857143\\
15.301 16.0571428571429\\
15.401 15.8285714285714\\
15.501 15.7142857142857\\
15.601 15.3142857142857\\
15.701 14.9714285714286\\
15.801 14.4571428571429\\
15.901 14.0571428571429\\
16.001 13.7142857142857\\
16.101 13.2571428571429\\
16.201 13.0285714285714\\
16.301 12.8571428571429\\
16.401 12.5714285714286\\
16.501 12.3428571428571\\
16.601 12.0571428571429\\
16.701 11.8857142857143\\
16.801 11.6571428571429\\
16.901 11.3142857142857\\
17.001 10.7428571428571\\
17.101 10.6285714285714\\
17.201 10.4571428571429\\
17.301 10.4571428571429\\
17.401 10.4571428571429\\
17.501 10.3428571428571\\
17.601 10.1142857142857\\
17.701 9.94285714285714\\
17.801 9.65714285714286\\
17.901 9.54285714285714\\
18.001 9.31428571428571\\
18.101 9.31428571428571\\
18.201 9.25714285714286\\
18.301 9.14285714285714\\
18.401 9.02857142857143\\
18.501 8.91428571428571\\
18.601 8.57142857142857\\
18.701 8.4\\
18.801 8.4\\
18.901 8.28571428571428\\
19.001 8.11428571428571\\
19.101 8.05714285714286\\
19.201 8\\
19.301 8\\
19.401 8\\
19.501 7.94285714285714\\
19.601 7.88571428571429\\
19.701 7.88571428571429\\
19.801 7.6\\
19.901 7.54285714285714\\
};
\node[right, inner sep=0mm, text=black]
at (axis cs:8.5,95.6571428571429,0) {$V_{100\%}$};
\node[right, inner sep=0mm, text=black]
at (axis cs:17.628,13,0) {$D_{10\%}$};
\addplot [
color=black,
dashed,
line width=3.0pt,
forget plot
]
table[row sep=crcr]{
0 93.6571428571429\\
8.5 93.6571428571429\\
};
\addplot [
color=black,
dashed,
line width=3.0pt,
forget plot
]
table[row sep=crcr]{
8.5 93.6571428571429\\
8.5 0\\
};
\addplot [
color=black,
dashed,
line width=3.0pt,
forget plot
]
table[row sep=crcr]{
17.628 0\\
17.628 10\\
};
\addplot [
color=black,
dashed,
line width=3.0pt,
forget plot
]
table[row sep=crcr]{
0 10\\
17.628 10\\
};
\addplot [
color=black,
mark size=1.7pt,
only marks,
mark=*,
mark options={solid},
forget plot
]
table[row sep=crcr]{
8.5 93.6571428571429\\
};
\addplot [
color=black,
mark size=1.7pt,
only marks,
mark=*,
mark options={solid},
forget plot
]
table[row sep=crcr]{
12.75 30.8571428571429\\
};
\addplot [
color=black,
mark size=1.7pt,
only marks,
mark=*,
mark options={solid},
forget plot
]
table[row sep=crcr]{
8.836 90\\
};
\end{axis}
\end{tikzpicture}%
\end{center}
\end{figure}

\emph{DVH criteria.} The dose distribution within the PTV or an OAR is described in a dose-volume histogram (DVH). Such a histogram displays the percentage of the structure receiving at least a certain dose. An example histogram is displayed in Figure \ref{fig:dvhexample}. Points on this histogram are denoted by $D_{x}$ or $V_{y}$, e.g., $D_{10\%}$ is the minimum dose received in the hottest 10\% of the structure under consideration and $V_{100\%}$ represents the percentage of the volume receiving 100\% of the prescribed dose. In a slight deviation from the above notation, $D_{\max}$ is the maximal dose received in the structure.  These statistics are currently the most important quantitative evaluation criteria \citep{Hoskin2013}. Besides these, the treatment planner inspects the isodose lines in order to avoid undesired hot spots and cold spots.

Table \ref{tab:DVHcriteria} displays clinically used DVH criteria from a local hospital.  In order to achieve tumor control, $V_{100\%}$ of the PTV needs to be at least $90\%$, i.e., at least $90\%$ of the PTV's volume need to receive at least $100\%$ of the prescribed dose. The limitation of complications is expressed in DVH constraints on the rectum and urethra. $D_{10\%}\leq 7.2 \unit{Gy}$ for the rectum imposes that $90\%$ of the rectum's tissue may not receive more than $7.2 \unit{Gy}$. A hard upper bound for tissue in the rectum is set by $D_{\max}\leq 8 \unit{Gy}$. For the urethra, $10 \unit{Gy}$ is the upper bound for $90\%$ of the tissue. The remaining $10\%$ are allowed to reach radiation levels of at most $10.6 \unit{Gy}$.
 \begin{table}
\center
\caption{DVH protocol for a prescribed dose of $8.5$ Gy for HDR prostate brachytherapy based on \citet{Hoskin2007}.}
\label{tab:DVHcriteria}
\begin{tabular}{ccc}
\toprule
PTV & Rectum & Urethra \\
\midrule
$V_{100\%}\geq 90\%$ & $D_{10\%} \leq 7.2 \unit{Gy}$    & $D_{10\%} \leq 10   \unit{Gy}$\\
                     & $D_{\max} \leq 8   \unit{Gy}$    & $D_{\max} \leq 10.6 \unit{Gy}$\\
\bottomrule
\end{tabular}

\end{table}

\emph{Dose measurement.} In order to compute DVH statistics, the PTV and the OARs are discretized into a grid of calculation points. The dose rates $\dot{d}_{ij}$ describe the amount of radiation emitted from dwell position $j\in J$ towards calculation point $i\in I$ per second. The dose at a calculation point $i$ is equal to the sum of radiation received from each dwell position $j$:
\begin{equation*}
\left(\dot{D}t\right)_{i}=\sum\limits_{j\in J}\dot{d}_{ij}t_{j},
\end{equation*}
where $t_{j}$ is the dwell time at dwell position $j$. Let $\dot{D}$ be the $|I|\times |J|$ matrix with all dose rates $\dot{d}_{ij}$. The matrix-vector product $\dot{D}t$ yields a vector that denotes the dose per calculation point.

\emph{Literature review.} Traditionally, inverse planning for HDR brachytherapy is dose-based: the treatment planner prescribes a lower bound $L_i$ and and an upper bound $U_i$ on the desired dose for each calculation point. Typically these bounds are the same for all calculation points within the same structure. At each calculation point $i$, the received dose $(\dot{D}t)_{i}$ is compared to its respective prescribed lower and upper bounds. If the dose lies outside the interval $[L_{i},U_{i}]$, a penalty is imposed that is linear or quadratic in the deviation. A treatment plan is optimal if it minimizes the sum of penalties over all calculation points.
Linear penalty functions have been successfully implemented in the commercially available algorithms Inverse Planning by Simulated Annealing (IPSA) \citep{Lessard2001} and Hybrid Inverse Treatment Planning and Optimization (HIPO) \citep{Karabis2005}. Quadratic penalty functions are used by \citet{Lahanas2003,lahanasQ2003,Milickovic2002}.

The disadvantage of dose-based penalty functions is that the resulting treatment plans may need a posteriori adjustments as they do not necessarily adhere to the DVH criteria. More recently, dose-volume based optimization methods have been developed that directly optimize on DVH criteria. \citet{Panchal2008} has formulated a Harmony Search heuristic. \citet{Siauw2011} have presented a Mixed Integer Linear Programming (MILP) formulation and developed the heuristic Inverse Planning by Integer Program (IPIP) to solve it. IPIP determines a solution that is feasible and near-optimal for MILP by solving two Linear Programs (LPs). \citet{Gorissen2013} have devised a formulation in which they directly optimize an MILP using specific solver settings. \citet{Belien2009} have introduced a hybrid approach based on simulated annealing and linear programming. Their objective is dose-based, while the delivered radiation to OARs is subject to dose-volume constraints. For the rectal constraint $D_{10\%} \leq 7.2 \unit{Gy}$, the linear constraint $(\dot{D}t)_i \leq 7.2 \unit{Gy}$ needs to hold for 90\% of the calculation points. The authors utilize simulated annealing to determine which calculation points are in the group of 90\%, and then use linear programming to determine the dwell times. According to \citet{Belien2009}, the algorithm yields results superior to running the integer program alone based on ten problem instances with $30$ minutes computation time each. The authors suggest the development of a pure SA algorithm without linear programming.

While our focus is on HDR brachytherapy, we briefly mention related research in Intensity Modulated Radiotherapy (IMRT). Instead of irradiating the PTV by bringing a radioactive source in close proximity to or inside the tumor, the patient is irradiated with radiation fields at different angles around the patient. For fixed beam angles, the optimization problem for determining the intensity map of each field is mathematically equivalent to the optimization problem we study. Historically, these research fields have not been connected, perhaps due to the different scale of the problems. For optimization in IMRT, we refer to \citet{ehrgott2010} and the extensive literature review by \citet{zaghian2014}. Simulated annealing is also applied to IMRT treatment planning, e.g., \citet{cho1998} optimize penalty functions to meet dose criteria.

\emph{Our contribution.} This paper follows up on \citet{Belien2009}'s suggestion to develop a pure simulated annealing (SA) heuristic to optimize on DVH statistics. We present DOPSA (DVH Optimization by Pure Simulated Annealing), a novel algorithm that solves the same MILP model as \citet{Gorissen2013} and \citet{Siauw2011}. In this model the objective is to maximize $V_{100\%}$ for the PTV, i.e., the volume of the PTV receiving at least the prescribed dose, while meeting DVH constraints on the OARs. Our method exploits the speed of matrix multiplication and uses a specialized generator for neighboring solutions. The heuristic has been tuned by conducting many trial-and-error experiments. The parameters have been selected using a metamodel, and have been validated with an out-of-sample analysis.

The advantage of SA over the existing dose-volume based models lies in its simple implementation independent of costly LP or MILP solvers. Since optimization is often offered as a separate module for treatment planning systems, clinics may choose cheaper dose-based optimization modules or keep using them if the price difference is too large, leaving the potential of dose-volume based methods unused. Furthermore, the results in Section \ref{sec:results} show that our heuristic has clear advantages over existing methods in terms of speed and objective value.

A short optimization time is of practical importance. In the clinical workflow \citep{balvert2015dwell}, two treatment plans are made while the patient is anesthetized and continuously monitored by an anesthesia care team. Moreover, the radiation oncologist is on standby while the treatment planner makes the plan. Shortening the time needed to create a treatment plan improves the clinical workflow and reduces the treatment costs.

The implementation details of DOPSA will be explained in the upcoming section, followed by a discussion of the results and a comparison with existing algorithms in Section \ref{sec:results}.

\section{Methods} \label{sec:implementation}

In this section we provide in-depth information about DOPSA. Its final implementation is the result of a long trial-and-error process where many ideas have been tested. Here, we restrict ourselves to a description of the final algorithm. For the selection procedure and parameter tuning based on trial-and-error and a metamodel we refer the reader to \citet{Deist2013}. There, values for the parameters (discussed below) have been determined by manual simulation and a linear metamodel has been fitted to fine-tune each parameter. However, the explanatory power of the model was low due to the presence of statistical noise. Note that only the data of Patient 1 has been employed in the parameter tuning. 

First, we describe the optimization model (Section \ref{sec:model}) and the main steps in the algorithm (Section \ref{sec:alg}). Then we focus on the two most important steps, which are the generation of neighborhood states (Section \ref{sec:stategeneration}) and checking feasibility (Section \ref{sec:feasibility}).

\subsection{Optimization model}\label{sec:model}
We use the same model as \citet{Gorissen2013} and \citet{Siauw2011} to allow a comparison with their results. This model is based on the DVH protocol from Table \ref{tab:DVHcriteria} with exception for the constraint on $V_{100\%}$. Instead of formulating a constraint $V_{100\%}\geq 90\%$, the model is cast to maximize $V_{100\%}$ under $D_{10\%}$ and $D_{\max}$ constraints on the OARs. Additionally, to ensure that the dose distribution conforms to the PTV and to avoid high doses outside the PTV in regions where there is no OAR to limit the dose in that region, we also consider an artificial structure that surrounds the prostate at 2 mm distance. This collection of tissue is henceforth denoted as the `shell', and a constraint is added to ensure that the maximum dose in this structure is below $8.5 \unit{Gy}$. Furthermore, a dwell time modulation restriction (DTMR) is added to avoid excessively high dwell times. The DTMR prohibits the dwell times of two adjacent dwell positions within the same catheter to differ more than a factor two. A similar restriction is found in other algorithms \citep{baltas2009}. The complete optimization model \citep{Gorissen2013} is
\begin{subequations}
\begin{align}
\max & 
\frac{1}{|I_{PTV}|} \sum\limits_{i\in I_{PTV}}v_{i} \label{eq:of}\\
\text{s.t.} & \left(\dot{D}t\right)_{i}\geq L_{i}v_{i} &\forall i \in  I_{PTV}\label{eq:conslptv}\\
&\left(\dot{D}t\right)_{i}\leq L_{i}+(U_{i}-L_{i})(1-v_{i}) & \forall i \in I_{R}\cup I_{U}\cup I_{S}\label{eq:consluoar}\\ 
& \sum\limits_{i\in I_{s}} v_{i}\geq \tau_{s}|I_{s}| & \forall s\in \left\{ R,U,S \right\}\label{eq:constauoar}\\
& t_{j_{1}}\leq \gamma t_{j_{2}} & \forall j_{1} \in J \quad j_{2} \in \Gamma(j_1)\label{eq:consdtmr}\\
& v_{i}\in\left\{ 0,1 \right\} & \forall i \in I \label{eq:consvintegral}\\
& t_{j} \in \left[ 0,\infty \right) & \forall j \in J.
\label{eq:constinterval}
\end{align}
\end{subequations}
The binary variable $v_{i}$ indicates whether calculation point $i$ meets a certain dose criterion:
for a calculation point in the PTV, i.e., $i \in I_{PTV}$, the corresponding $v_{i}$ is equal to one if the received dose at those points is at least the prescribed dose of $L_{i}=8.5 \unit{Gy}$. These values are enforced by the objective \eqref{eq:of} and constraint \eqref{eq:conslptv}. For a calculation point in the OARs rectum (R), urethra (U), and shell (S), i.e., $i \in I_{R}\cup I_{U}\cup I_{S}$, constraint \eqref{eq:consluoar} allows $v_{i}$ to be one if the received dose is at most the soft upper bound $L_{i}$. If the received dose is between $L_{i}$ and the hard upper bound $U_{i}$, $v_{i} = 0$. Constraint \eqref{eq:constauoar} enforces the soft upper bound on the dose for at least a fraction $\tau_{s}$ of the calculation points. The corresponding parameter values are displayed in Table \ref{tab:DVHbounds}. Note that for the shell, $L_{i}$ and $U_{i}$ coincide and $\tau_{s} = 1$, since only a single dose constraint is imposed for this structure.
 Constraint \eqref{eq:consdtmr} enforces the DTMR, i.e., dwell times at neighboring dwell positions in the same catheter may not differ by more than a factor $\gamma$. For our implementation, $\gamma = 2$. The set $\Gamma(j)$ contains all adjacent dwell positions of $j$. 

 \begin{table}
\center
\caption{Radiation bounds and $\tau$ for each structure.}
\begin{tabular}{ccccc}
\toprule & PTV & Rectum & Urethra & Shell \\
\midrule
$L_{i}$ & $8.5 \unit{Gy}$  & $7.2 \unit{Gy}$  & $10 \unit{Gy}$ & $8.5 \unit{Gy}$ \\
$U_{i}$ & n/a & $8 \unit{Gy}$  & $10.6 \unit{Gy}$  & $8.5 \unit{Gy}$ \\
$\tau_{s}$ & n/a & $0.9$ & $0.9$ & $1$\\
\bottomrule
\end{tabular}
\label{tab:DVHbounds}
\end{table}

\subsection{Algorithm}\label{sec:alg}
The algorithm described below searches for a vector $t$ such that objective \eqref{eq:of} is maximal and all constraints \eqref{eq:conslptv}-\eqref{eq:constinterval} are satisfied.
DOPSA has been implemented in MATLAB according to Algorithm \ref{alg:SAsimple}, an adapted version of a standard simulated annealing process \citep{kirkpatrick}. In simulated annealing, the feasible region is searched by selecting a new solution $t^{new}$ in the neighborhood of the current solution $t^{cur}$ and deciding whether to accept this new choice. A solution which is not decreasing the objective function is always accepted, whereas decreasing solutions are also accepted with a probability that decreases in the loss of objective function value and the duration of the search. This rule is also named the metropolis criterion. The acceptance probability is defined as
\begin{equation*} \mathbb{P} \left(\text{accept $t^{new}$}\right) = \min \left\{ 1,\exp\tfrac{f\left(t^{new}\right)-f\left(t^{cur}\right)}{c_{k}}
\right\}
\end{equation*}
 with objective function $f$ and $c_{k}$ regulating the probability's sensitivity to lower objective function values. The parameter $c_{k}$ is adjusted at each iteration $k$ according to the well-known exponential cooling schedule
 \begin{equation*}
  c_{k}=c_{0}\alpha^{\left\lfloor\frac{k}{m}\right\rfloor},
\end{equation*}
where  $m\in(0,\infty)$ is a fixed number of iterations after which the initial temperature $c_{0}$ is rescaled by $\alpha\in (0,1)$.

DOPSA's code is designed to generate and evaluate multiple states simultaneously. All time-consuming procedures within the heuristic thus become matrix-operations which decreases the required CPU time per state. Essentially it is faster to check the feasibility of $g$ states in one operation than to check the feasibility of the $g$ states individually, since computing $\dot{D}T$, where $T$ is a $|J| \times g$ matrix with the $g$ states as columns, takes less than $g$ times longer than computing $\dot{D}t$ for a single state $t$. This not only eliminates a for-loop in Matlab code, which is known to be relatively slow, but more importantly, uses the fact that an efficient implementation of a matrix multiplication procedure takes less operations than $\mathcal{O}(g|I||J|)$ when $g \geq 2$.

\begin{algorithm}
\caption{Simulated Annealing Code}
\label{alg:SAsimple}
\begin{algorithmic}
\State set the initial state to $t = 0$
\State set the temperature to $c_{0}$
\While{running time $\leq$ 180 seconds}
\State every 15000 iterations: return to current optimum
\State generate neighborhood states
\State{discard infeasible states}
\State{choose $t^{new}$ from neighborhood according to highest $V_{100\%}$}
\State{decide acceptance}
\State lower the temperature by scaling with $\alpha$
\EndWhile
\end{algorithmic}
\end{algorithm}

DOPSA is initialized with the dwell times set to zero because it should start in the feasible region and, a priori, any other feasible state can only be determined at relatively high computational effort. Then, one state is chosen from the set of feasible candidates in the neighborhood and is evaluated by the metropolis criterion as described in Section \ref{sec:implementation}. Subsequently, the temperature is decreased in every iteration according to an exponential cooling schedule with parameters $\alpha=0.99$, $m=1$, and $c_{0}=1.5$. The search continues in the chosen state, if accepted, or restarts in the current state. After each 15,000 iterations, the state with the highest $V_{100\%}$ is selected and the search is restarted in the neighborhood of this state. This step is included to recover from an unsuccessful attempt to escape from a local optimum. Returning to the state with the highest $V_{100\%}$ focuses the search to areas with high objective function values and avoids unnecessary search efforts in low-potential regions.

The running time of DOPSA is restricted to a fixed limit (e.g., 180 seconds) to allow a comparison with existing methods and to test whether DOPSA can provide satisfactory results within a short time-span. In practice, an alternative stopping criterion can be used, such as the relative improvement of the objective value over the last $n$ iterations, or user intervention based on the coverage attained by the algorithm and the running time.

The performance of the algorithm was revealed to be insensitive to changes in the parameter values of $\alpha$ and $c_{0}$, suggesting that hill-climbing methods might be similarly successful.

\subsection{Generating neighborhood states}\label{sec:stategeneration}
DOPSA generates a neighborhood solution $t$ by perturbing the dwell times of the current state $t^{cur}$:
\begin{equation*}
\label{eq:newt}
t=t^{cur}+\left\lfloor r\right\rceil,
\end{equation*}
where $r$ is a vector of the same size as $t$. The vector $r$ is chosen randomly according to $r_{j}\sim N(0,0.05)$ i.i.d. for all $j$ in $J_{c}$, the set of dwell positions that are changed, and $r_{j}=0$ for all $j$ in $J\backslash J_{c}$. The operator $\lfloor\rceil$ rounds the elements of $r$ to one decimal place, which is the input precision of the Flexitron remote afterloader (Nucletron BV, Veenendaal, the Netherlands). Negative dwell times are subsequently set to 0.

For the DTMR, the dwell times in each catheter are adjusted one by one to ensure the relative difference with the previous dwell time is less than a factor 2. This adjustment may increase the number of changed entries in the dwell time vector.

\emph{Number of changed entries.} We observed that controlling the number of dwell time changes applied to a new state is fundamental to the performance of DOPSA. A lower number of changes is more likely to yield a feasible new state. However, lowering this number decreases the potential maximal increase in $V_{100\%}$ in the new state, which is most notable in early iterations. Since it is unlikely that big improvements can still be realized when the algorithm stalls, a negative correlation between the number of changed entries and the improvement over the past 200 iterations is introduced. Consequently, when the improvements get smaller, newly generated states get a larger probability of being feasible while the risk of missing out on potential large gains is minimal. In the first 200 iterations, all entries are subject to changes before the dynamic adjustment begins. The warmup period of 200 iterations only affects the initial phase of DOPSA since the total number of iterations is approximately 50000.

Let $J_{c}$ denote the index set of entries in $t^{cur}$ that are varied to construct a new vector $t$. The relation between the cardinality of $J_c$ and the improvement in $V_{100\%}$ is defined by
\begin{align*}
|J_{c}| =\left\lceil \tfrac{V_{100\%}^{k}-V_{100\%}^{k-200,max}}{V_{100\%}^{k-200,max}}|J|\right\rceil, \label{eq:Jc}
\end{align*}
where $V_{100\%}^{k}$ is the $V_{100\%}$ attained in the current state $t^{cur}$ at iteration $k$. $V_{100\%}^{k-200,max}$ is the overall highest $V_{100\%}$ attained until iteration $k-200$.

Note that the states chosen during the search can also exhibit $V_{100\%}$ lower than in preceding iterations due to the metropolis criterion, resulting in a negative $|J_c|$. For that reason and to maintain a natural upper bound, $|J_{c}|$ is restricted to the interval $[\lceil 0.02 |J|\rceil,|J|]$. The lower bound is chosen to be at least 1 (since $|J| > 50$ for practically all cases) since improvements in $V_{100\%}$ might only be attained with increases in dwell times for one dwell position while simultaneously decreasing the dwell time of another position. Taking $2\%$ of all entries as the lower bound ensures sufficient dwell time changes in generated states.

Empirical results indicate that a newly generated state has six times the probability of being feasible by using $J_c$ compared to perturbing all dwell times. The number of states that are dropped due to infeasibility decreases from 95\% to 70\%.

\emph{Dwell position selection.} Once $|J_c|$, the number of dwell positions for which the dwell times are perturbed, is chosen, the next step is to select that many dwell positions. The intention is to alter dwell times such that the objective value is improved. This target is pursued by changing the dwell times of the dwell positions close to the calculation points in the PTV that do not yet receive the prescribed dose. Since the distance between dwell position $j$ and calculation point $i$ has an inverse squared relation to the dose rate $\dot{d}_{ij}$, the probability that dwell position $j$ is chosen in a single draw is therefore set to
 \begin{equation*}\label{eq:dwellpositionprobability}
P\left(j\in J_{c}\right)=\tfrac{\sum\limits_{i\in I_{PTV} : (\dot{D}t)_{i}<8.5}\dot{d}_{ij}}{\sum\limits_{\tilde{j}\in J}\sum\limits_{i\in I_{PTV} : (\dot{D}t)_{i}<8.5}\dot{d}_{i\tilde{j}} },
\end{equation*}
where $I_{PTV}$ is the set of calculation points in the PTV. The denominator normalizes the term to attain a probability distribution. Using probabilities rather than simply picking the closest dwell positions bears the advantage that the set of states that can be generated from the current state does not become too small. Occasionally also dwell positions further away are selected which diversifies the search.

This methods requires a random sample of dwell positions in each iteration of the search. Randomly sampling dwell positions without replacement in MATLAB is a computationally expensive procedure. Our experiments have shown that sampling $|J_c|$ items with replacement and removing duplicates yields better results, due to a larger number of iterations in the same running time. In a direct comparison, the average PTV coverage after a running time of 3 minutes over 50 replications was 0.12\% higher when using sampling without replacement.

\emph{Number of generated states.} In the early stages of the search process, improving states can be determined with few generated states per iteration since the improvement potential in the neighborhood of $t^{cur}$ is high. Later on, an intensification of the search in those neighborhoods is most important. Moreover, intensification becomes necessary only when the search approaches the boundary, where one expects to find optimal solutions. Therefore, DOPSA generates more states when the boundary is reached (as suggested by \citet{HedarFukushima2006}). Intensification can be controlled with an opposite relation between the number of states generated in a single iteration $g$ and $V_{100\%}$:
\begin{align*}
g = \left\lceil 40 \left( 1-\tfrac{V_{100\%}^{k}-V_{100\%}^{k-200,max}}{V_{100\%}^{k-200,max}} \right) \right\rceil. \label{eq:g}
\end{align*}
$g$ is limited to the interval $[1,40]$ because trial runs indicated that excessive state generation does not further improve the DOPSA's performance. In the first 200 iterations, $g=1$ is used.

\subsection{Checking feasibility}\label{sec:feasibility}
The feasibility of each generated state needs to be verified, which is one of the computationally most expensive procedures next to computing $V_{100\%}$. Testing each constraint discussed in Section \ref{sec:introduction} requires a row subset of the matrix $\dot{D}$ corresponding to a specific OAR to be multiplied with the dwell time vector $t$. The computation time of checking a constraint increases with the number of calculation points in the corresponding OAR.

The sequence in which the constraints on the DVHs for the OARs are checked is chosen to minimize the required computation time: the most restrictive constraints with lowest computational cost are evaluated first. States that do not satisfy a constraint are immediately excluded in order to reduce the computation time for checking other constraints.

The calculation of constraint values and $V_{100\%}$ for newly generated states can be sped up by using the values for the current state $t^{cur}$ that have been computed in the preceding iteration. Let $T^{cur}$ denote the $|J|\times g$ matrix containing the current state as columns, and let $T^{new}$ denote the $|J|\times g$ matrix with the neighboring states as columns.

The dose computation for each generated state $\dot{D}T^{new}$ can be reformulated as follows:
\begin{equation*}
\dot{D}T^{new}  =  \dot{D}T^{cur}+\dot{D}\left(T^{new} - T^{cur}\right). \label{eq:sparsity1}
\end{equation*}
The matrix product $\dot{D}T^{cur}$ is known from the previous iteration, so only $\dot{D}(T^{new} - T^{cur})$ needs to be computed. $T^{new} - T^{cur}$ is a sparse matrix with at most $|J_c|$ non-zero entries per column (plus additional non-zero entries due to the DTMR adjustment), since states generated in the neighborhood of a current state only differ in approximately $|J_{c}|$-many entries. Moreover, the sparsity increases over the search process. Using MATLAB's toolpack for multiplication of sparse matrices, an increase of the number of iterations over the total running time of 5\% up to 21\% could be observed, depending on the number of dwell positions.

\section{Results}
\label{sec:results}
The performance of DOPSA is assessed by a direct comparison with two existing dose-volume-based methods: the MILP formulation by \citet{Gorissen2013} and the IPIP algorithm by \citet{Siauw2011}. The running time has been fixed to 3 minutes on an Intel Xeon E5620 2.4 GHz with 6 GB of memory. Both MILP and IPIP are solved using the state-of-the-art solver CPLEX 12.4 (IBM).

\subsection{Patient data}
Tables \ref{tab:datasetcalcpoints} and \ref{tab:datasetdwellpos} provide an overview of the three patient data sets. Each data set uses the same number of calculation points in the PTV and the OARs. The PTV contains the biggest share of calculation points since it is the largest organ. All dwell positions inside the PTV have been activated based on a step size of 2.5 mm, resulting in 115 to 236 dwell positions per patient. The number of inserted catheters is similar among patients with 17 for Patient 2 and 16 for the other two patients. The dose rates have been calculated using the TG-43 formalism.

The calculation points have been generated with a Sobol sequence, which is a quasi-random sampling technique that randomly distributes calculation points in such a way that the sampling density is approximately uniform. This avoids the possibility that small subregions are not covered with calculation points. Quasi-random sampling was shown to be excellent for evaluating treatment plans \citep{niemierko1990random}, although this effect may disappear when the treatment plan is optimized with respect to the randomly sampled calculation points. To avoid excessively high doses in points close to dwell positions, distances smaller than 0.83 mm are set to 0.83 mm (in the same direction).

\begin{table}[htbp]
  \centering
  \caption{The number of calculation points per structure.}
    \begin{tabular}{lr}
    \toprule
    Structure & Calculation Points \\
    \midrule
    PTV     & 1750 \\
    Urethra &  500 \\
    Rectum  &  250 \\
    Shell   &  250 \\
    \bottomrule
    \end{tabular}%
  \label{tab:datasetcalcpoints}%
\end{table}%

\begin{table}[htbp]
  \centering
  \caption{The number of dwell positions and catheters for each patient.}
    \begin{tabular}{rrr}
    \toprule
    Patient & Dwell Positions & Catheters \\
    \midrule
    1     & 236   & 16 \\
    2     & 115   & 17 \\
    3     & 182   & 16 \\
    \bottomrule
    \end{tabular}
  \label{tab:datasetdwellpos}
\end{table}

\subsection{Running time and objective value}
DOPSA has been tested on three patients. Since it is a random local search procedure, the result can be subject to random effects. Therefore, the heuristic has been run 250 times for each patient in order to provide a detailed analysis of the average treatment plan quality and the variation in quality. If a low variation can be ensured for this optimization algorithm, only a single optimization run per patient will be necessary in a clinical setting to obtain treatment plans with consistent quality.

\begin{figure}
\caption{Average and standard deviation in PTV coverage for Patient 1 (250 replications).}
\label{fig:sapt2}
\begin{center}
\input{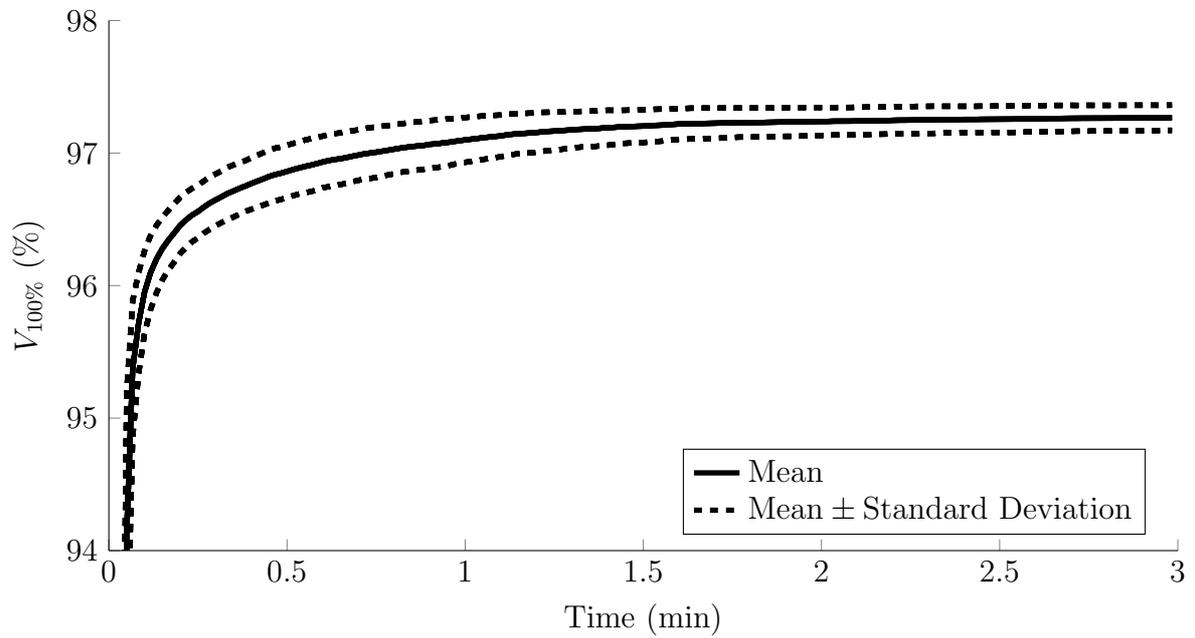}
\end{center}
\end{figure}

Figures \ref{fig:sapt2}-\ref{fig:sapt4} show the results for all three patient data sets. The average and standard deviation in PTV coverage over 250 replications are displayed over the running time of 180 seconds.  The positioning of catheters determines the maximally achievable level of $V_{100\%}$ in the PTV, which explains the difference in objective values between patients. The high dependence on catheter positioning is well observable for Patient 2 for whom only solutions with $V_{100\%}<50\%$ can be attained: inspecting the corresponding ultrasound scans reveals badly positioned catheters.

In all three cases, the PTV coverage increases rapidly within the first 30 seconds and further improvements on the treatment plan quality are achieved over the remaining running time. Every treatment plan determined by DOPSA is feasible, i.e., it always satisfies the DVH criteria imposed on the OARs and dwell time constraints (see constraints \eqref{eq:consluoar}-\eqref{eq:constinterval}).

The absolute standard deviations in PTV coverage after a running time of 3 minutes for each of the 250 replications are 0.09\%, 0.33\%, and 0.23\%, respectively. Therefore, DOPSA delivers treatment plans with consistent quality across replications.

\begin{figure}
\caption{Average and standard deviation in PTV coverage for Patient 2 (250 replications).}
\label{fig:sapt3}
\begin{center}
\input{sapt3.tikz}
\end{center}
\end{figure}

\begin{figure}
\caption{Average and standard deviation in PTV coverage for Patient 3 (250 replications).}
\label{fig:sapt4}
\begin{center}
\input{sapt4.tikz}
\end{center}
\end{figure}

\subsection{Comparison with existing DVH-based optimizers}
\label{sec:comparison}
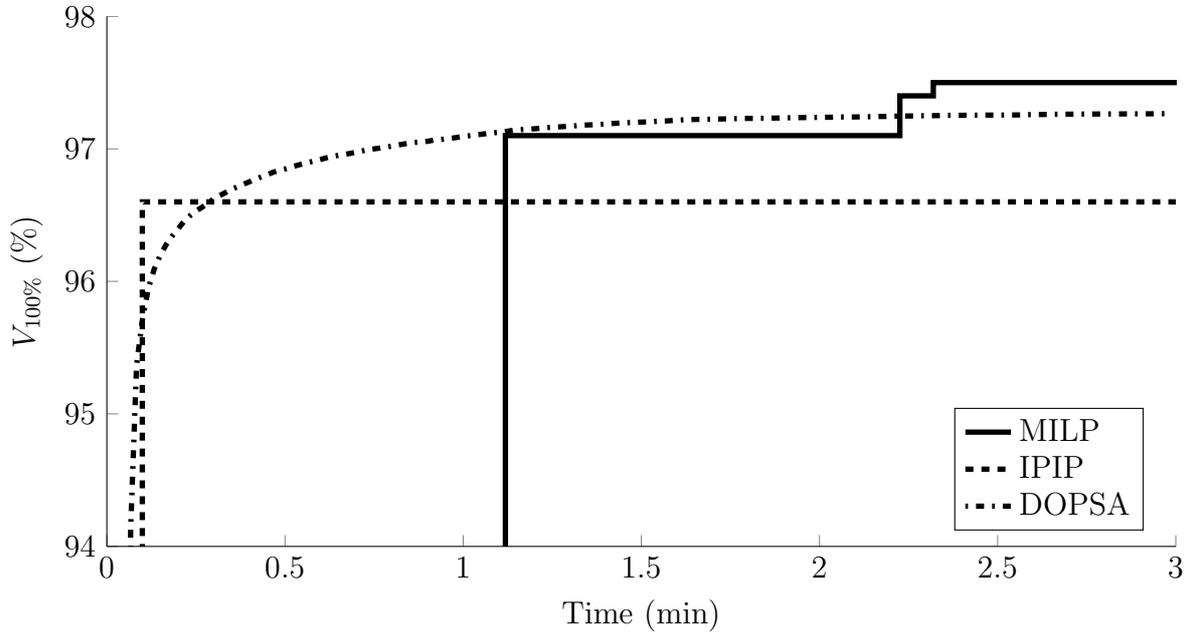
\begin{figure}
\caption{Comparison of PTV coverage for DVH-based optimizers for Patient 1.}
\label{fig:DVHpt2}
\begin{center}
% This file was created by matlab2tikz v0.4.0.
% Copyright (c) 2008--2013, Nico Schlömer <nico.schloemer@gmail.com>
% All rights reserved.
% 
% The latest updates can be retrieved from
%   http://www.mathworks.com/matlabcentral/fileexchange/22022-matlab2tikz
% where you can also make suggestions and rate matlab2tikz.
% 
% 
% 
\begin{tikzpicture}

\begin{axis}[%
width=400pt,
height=200pt,
scale only axis,
xmin=0,
xmax=180,
xtick={0,30,60,90,120,150,180},
xticklabels={0,0.5,1,1.5,2,2.5,3},
xlabel={Time (min)},
ymin=94,
ymax=98,
ytick={ 90,  91,  92,  93,  94,  95,  96,  97,  98,  99, 100},
ylabel={$V_{100\%}$ (\%)},
axis x line*=bottom,
axis y line*=left,
legend style={at={(0.97,0.03)},anchor=south east,draw=black,fill=white,legend cell align=left}
]
\addplot [
color=black,
solid,
line width=2.0pt
]
table[row sep=crcr]{
67.11 0\\
67.11 97.1\\
67.21 97.1\\
67.21 97.1\\
67.46 97.1\\
67.46 97.1\\
67.91 97.1\\
67.91 97.1\\
67.92 97.1\\
67.92 97.1\\
68.06 97.1\\
68.06 97.1\\
71 97.1\\
71 97.1\\
71.34 97.1\\
71.34 97.1\\
76.04 97.1\\
76.04 97.1\\
77.19 97.1\\
77.19 97.1\\
79.98 97.1\\
79.98 97.1\\
80.12 97.1\\
80.12 97.1\\
82.4 97.1\\
82.4 97.1\\
83.59 97.1\\
83.59 97.1\\
86.19 97.1\\
86.19 97.1\\
86.5 97.1\\
86.5 97.1\\
86.69 97.1\\
86.69 97.1\\
89.03 97.1\\
89.03 97.1\\
89.25 97.1\\
89.25 97.1\\
93.07 97.1\\
93.07 97.1\\
95.47 97.1\\
95.47 97.1\\
96.8 97.1\\
96.8 97.1\\
96.89 97.1\\
96.89 97.1\\
97.08 97.1\\
97.08 97.1\\
99.58 97.1\\
99.58 97.1\\
99.95 97.1\\
99.95 97.1\\
100.59 97.1\\
100.59 97.1\\
100.73 97.1\\
100.73 97.1\\
101.14 97.1\\
101.14 97.1\\
101.28 97.1\\
101.28 97.1\\
106.83 97.1\\
106.83 97.1\\
107.19 97.1\\
107.19 97.1\\
107.81 97.1\\
107.81 97.1\\
108.19 97.1\\
108.19 97.1\\
111.98 97.1\\
111.98 97.1\\
112.13 97.1\\
112.13 97.1\\
112.26 97.1\\
112.26 97.1\\
114.04 97.1\\
114.04 97.1\\
114.18 97.1\\
114.18 97.1\\
115.19 97.1\\
115.19 97.1\\
115.64 97.1\\
115.64 97.1\\
115.88 97.1\\
115.88 97.1\\
119.7 97.1\\
119.7 97.1\\
120.18 97.1\\
120.18 97.1\\
120.92 97.1\\
120.92 97.1\\
121.1 97.1\\
121.1 97.1\\
121.31 97.1\\
121.31 97.1\\
121.42 97.1\\
121.42 97.1\\
121.98 97.1\\
121.98 97.1\\
122.13 97.1\\
122.13 97.1\\
124.43 97.1\\
124.43 97.1\\
124.55 97.1\\
124.55 97.1\\
125.74 97.1\\
125.74 97.1\\
125.88 97.1\\
125.88 97.1\\
128.62 97.1\\
128.62 97.1\\
128.76 97.1\\
128.76 97.1\\
128.97 97.1\\
128.97 97.1\\
130.78 97.1\\
130.78 97.1\\
130.85 97.1\\
130.85 97.1\\
130.89 97.1\\
130.89 97.1\\
131.34 97.1\\
131.34 97.1\\
133.54 97.1\\
133.54 97.4\\
133.69 97.4\\
133.69 97.4\\
133.99 97.4\\
133.99 97.4\\
134.18 97.4\\
134.18 97.4\\
134.27 97.4\\
134.27 97.4\\
134.54 97.4\\
134.54 97.4\\
134.85 97.4\\
134.85 97.4\\
135.08 97.4\\
135.08 97.4\\
135.18 97.4\\
135.18 97.4\\
136.11 97.4\\
136.11 97.4\\
136.38 97.4\\
136.38 97.4\\
136.46 97.4\\
136.46 97.4\\
136.55 97.4\\
136.55 97.4\\
139.17 97.4\\
139.17 97.5\\
139.4 97.5\\
139.4 97.5\\
139.78 97.5\\
139.78 97.5\\
140.11 97.5\\
140.11 97.5\\
140.57 97.5\\
140.57 97.5\\
140.67 97.5\\
140.67 97.5\\
140.93 97.5\\
140.93 97.5\\
140.95 97.5\\
140.95 97.5\\
140.96 97.5\\
140.96 97.5\\
140.98 97.5\\
140.98 97.5\\
141.14 97.5\\
141.14 97.5\\
141.14 97.5\\
141.14 97.5\\
141.32 97.5\\
141.32 97.5\\
141.59 97.5\\
141.59 97.5\\
141.67 97.5\\
141.67 97.5\\
141.68 97.5\\
141.68 97.5\\
143.05 97.5\\
143.05 97.5\\
143.26 97.5\\
143.26 97.5\\
143.35 97.5\\
143.35 97.5\\
143.51 97.5\\
143.51 97.5\\
143.58 97.5\\
143.58 97.5\\
144.4 97.5\\
144.4 97.5\\
145.13 97.5\\
145.13 97.5\\
145.61 97.5\\
145.61 97.5\\
145.97 97.5\\
145.97 97.5\\
146.24 97.5\\
146.24 97.5\\
147.03 97.5\\
147.03 97.5\\
149.78 97.5\\
149.78 97.5\\
149.98 97.5\\
149.98 97.5\\
150.7 97.5\\
150.7 97.5\\
152.82 97.5\\
152.82 97.5\\
153.44 97.5\\
153.44 97.5\\
153.93 97.5\\
153.93 97.5\\
155.6 97.5\\
155.6 97.5\\
155.77 97.5\\
155.77 97.5\\
158.22 97.5\\
158.22 97.5\\
159.75 97.5\\
159.75 97.5\\
159.93 97.5\\
159.93 97.5\\
163.15 97.5\\
163.15 97.5\\
163.46 97.5\\
163.46 97.5\\
165.91 97.5\\
165.91 97.5\\
166.24 97.5\\
166.24 97.5\\
168.37 97.5\\
168.37 97.5\\
168.45 97.5\\
168.45 97.5\\
171.06 97.5\\
171.06 97.5\\
171.41 97.5\\
171.41 97.5\\
174.91 97.5\\
174.91 97.5\\
175.17 97.5\\
175.17 97.5\\
178.12 97.5\\
178.12 97.5\\
178.26 97.5\\
178.26 97.5\\
178.37 97.5\\
178.37 97.5\\
178.5 97.5\\
178.5 97.5\\
180.98 97.5\\
};
\addlegendentry{MILP};

\addplot [
color=black,
dashed,
line width=2.0pt
]
table[row sep=crcr]{
5.97 0\\
5.97 96.6\\
180 96.6\\
};
\addlegendentry{IPIP};

\addplot [
color=black,
dash pattern=on 1pt off 3pt on 3pt off 3pt,
line width=2.0pt
]
table[row sep=crcr]{
3 88.3865142857143\\
4 94.1295999999998\\
5 95.3693714285713\\
6 95.7156571428571\\
7 95.9497142857142\\
8 96.0955428571428\\
9 96.1981714285713\\
10 96.2781714285713\\
11 96.3414857142856\\
12 96.3963428571427\\
13 96.4532571428571\\
14 96.4953142857141\\
15 96.5311999999999\\
16 96.5581714285713\\
17 96.5917714285713\\
18 96.6223999999998\\
19 96.6470857142856\\
20 96.6706285714285\\
21 96.6950857142856\\
22 96.7140571428571\\
23 96.7337142857142\\
24 96.7513142857142\\
25 96.7702857142856\\
26 96.7881142857142\\
27 96.8050285714284\\
28 96.8235428571427\\
29 96.8374857142856\\
30 96.8482285714285\\
31 96.8612571428571\\
32 96.8751999999999\\
33 96.8870857142856\\
34 96.8987428571428\\
35 96.9085714285713\\
36 96.9199999999999\\
37 96.9325714285713\\
38 96.9430857142856\\
39 96.9510857142856\\
40 96.9574857142856\\
41 96.9659428571428\\
42 96.9757714285713\\
43 96.9846857142856\\
44 96.994057142857\\
45 96.9999999999999\\
46 97.0061714285713\\
47 97.0137142857142\\
48 97.0210285714285\\
49 97.0287999999999\\
50 97.0365714285713\\
51 97.0429714285714\\
52 97.0486857142856\\
53 97.053257142857\\
54 97.0571428571427\\
55 97.0655999999999\\
56 97.0687999999999\\
57 97.0747428571428\\
58 97.0799999999999\\
59 97.0863999999999\\
60 97.0932571428571\\
61 97.0989714285714\\
62 97.1044571428571\\
63 97.1097142857142\\
64 97.115657142857\\
65 97.120457142857\\
66 97.1250285714285\\
67 97.1293714285713\\
68 97.1371428571427\\
69 97.1421714285713\\
70 97.1458285714284\\
71 97.1485714285713\\
72 97.1515428571427\\
73 97.1549714285713\\
74 97.1599999999998\\
75 97.1634285714284\\
76 97.1657142857141\\
77 97.1682285714284\\
78 97.1721142857141\\
79 97.1743999999998\\
80 97.1785142857141\\
81 97.1805714285713\\
82 97.1821714285713\\
83 97.1842285714284\\
84 97.1878857142855\\
85 97.1910857142855\\
86 97.1935999999998\\
87 97.1970285714284\\
88 97.1983999999998\\
89 97.1999999999998\\
90 97.202057142857\\
91 97.2031999999998\\
92 97.2057142857141\\
93 97.2086857142855\\
94 97.2102857142855\\
95 97.2130285714284\\
96 97.2162285714284\\
97 97.2187428571427\\
98 97.2203428571427\\
99 97.2221714285712\\
100 97.2226285714284\\
101 97.2230857142855\\
102 97.2242285714284\\
103 97.2251428571427\\
104 97.2260571428569\\
105 97.2269714285712\\
106 97.2285714285712\\
107 97.2297142857141\\
108 97.2299428571427\\
109 97.2299428571427\\
110 97.2306285714284\\
111 97.2308571428569\\
112 97.2319999999998\\
113 97.2326857142855\\
114 97.2335999999998\\
115 97.2347428571427\\
116 97.2349714285712\\
117 97.2361142857141\\
118 97.2361142857141\\
119 97.2370285714284\\
120 97.2379428571427\\
121 97.2383999999998\\
122 97.2386285714284\\
123 97.2402285714284\\
124 97.2404571428569\\
125 97.2404571428569\\
126 97.2406857142855\\
127 97.2420571428569\\
128 97.2422857142855\\
129 97.2431999999998\\
130 97.2443428571427\\
131 97.2445714285712\\
132 97.2454857142855\\
133 97.246857142857\\
134 97.2477714285712\\
135 97.2479999999998\\
136 97.248457142857\\
137 97.2495999999998\\
138 97.2502857142855\\
139 97.2505142857141\\
140 97.2509714285712\\
141 97.2523428571427\\
142 97.2527999999998\\
143 97.2530285714284\\
144 97.2530285714284\\
145 97.2539428571427\\
146 97.2539428571427\\
147 97.2539428571427\\
148 97.2546285714284\\
149 97.254857142857\\
150 97.2550857142855\\
151 97.2553142857141\\
152 97.2555428571427\\
153 97.2557714285713\\
154 97.2564571428569\\
155 97.2578285714284\\
156 97.2587428571427\\
157 97.2591999999998\\
158 97.2594285714284\\
159 97.2607999999998\\
160 97.2610285714284\\
161 97.2610285714284\\
162 97.2617142857141\\
163 97.2621714285712\\
164 97.2623999999998\\
165 97.2628571428569\\
166 97.2628571428569\\
167 97.2630857142855\\
168 97.2630857142855\\
169 97.2633142857141\\
170 97.2633142857141\\
171 97.2637714285712\\
172 97.2639999999998\\
173 97.2642285714284\\
174 97.2642285714284\\
175 97.2644571428569\\
176 97.2646857142855\\
177 97.2651428571426\\
178 97.2655999999998\\
179 97.2660571428569\\
};
\addlegendentry{DOPSA};

\end{axis}
\end{tikzpicture}%
\end{center}
\end{figure}

For the comparison with existing DVH-based optimizers, we have run an MILP solver with the solver settings from \citet{Gorissen2013} and we have implemented IPIP \citep{Siauw2011}. The original IPIP has been extended by the DTMR to ensure a fair comparison. These two methods have been run on the same three patient data sets and on the same model. It is important to recognize that IPIP can only provide a single solution per data set, whereas MILP and DOPSA continue improving the solution. This is observable in Figures \ref{fig:DVHpt2}-\ref{fig:DVHpt4}, where the line for IPIP remains flat after a single increase. In theory, MILP could stop when it proves that the solution is optimal, but this did not happen for any of the patients.

Figures \ref{fig:DVHpt2} to \ref{fig:DVHpt4} show the average PTV coverage obtained by DOPSA and the coverage obtained by the existing methods MILP and IPIP. The running time for Patient 3 has been extended to 5 minutes since MILP requires more than 3 minutes to find non-zero feasible points.

IPIP attains adequate PTV coverage within less than 6 seconds for each patient. All those solutions are initially superior to treatment plans found by DOPSA. Over the running time of 3 minutes, however, DOPSA consistently outperforms IPIP for all patient data sets. In each of the 250 replications for all three patients, the PTV coverage after 3 minutes is higher for DOPSA than for IPIP. The average PTV coverage attained by DOPSA exceeds the results by IPIP after only 18, 6, and 12 seconds, respectively.

MILP requires substantially more time to determine good treatment plans. For 2 out of 3 patients, MILP requires more than 1 minute to determine a treatment plan with non-zero PTV coverage. However, after 3 minutes for Patients 1-2 and 5 minutes for Patient 3 it has found the best plan among the three algorithms.

Unfortunately, the optimal objective values are not known. There is still a large gap between the best solution found by any of the algorithms and the upper bound provided by the MILP solver, and this gap does not decrease substantially after running the solver for a week. DOPSA differs greatly from the existing dose-volume based methods while it finds similar objective values. This could be an indication that much better solutions do not exist.

\begin{figure}
\caption{Comparison of PTV coverage for DVH-based optimizers for Patient 2.}
\label{fig:DVHpt3}
\begin{center}
\input{allDVHpt3.tikz}
\end{center}
\end{figure}
\begin{figure}
\caption{Comparison of PTV coverage for DVH-based optimizers for Patient 3.}
\label{fig:DVHpt4}
\begin{center}
\input{allDVHpt4.tikz}
\end{center}
\end{figure}

\subsection{Model Discussion}

The significance of the strong performance of DOPSA compared to the MILP by \citet{Gorissen2013} becomes apparent when closely investigating the structure of the MILP \eqref{eq:of}-\eqref{eq:constinterval}. Below, we show that the MILP is equivalent to the well-known $\ell_{0}$-norm minimization, and that an MILP solver implicitly uses a method that is known to be effective in approximating the $\ell_{0}$-norm. Therefore, we conclude that DOPSA displays a remarkable performance by outperforming MILP in terms of solution speed. 

The objective \eqref{eq:of} is to maximize the number of calculation points that receive the prescribed dose of 8.5 Gy. Let us first define the vector $y$, indexed by $i$ in $I_{PTV}$ as follows:
\begin{equation*}
y_{i}=\max\left\{0, \tfrac{8.5 - \left(\dot{D}t\right)_{i}}{8.5} \right\}.
\end{equation*}
Since $y_{i}$ is zero if calculation point $i$ receives at least the prescribed dose, minimizing the number of non-zero elements in $y$ yields a treatment plan with maximal $V_{100\%}$. The number of non-zero elements is $||y||_{0}$ (recall that the $\ell_{0}$-norm counts the number of non-zero elements in a vector). $D_{10\%}$ and $D_{\max}$ can be rewritten similarly. Thus, the Lagrangean of the MILP \eqref{eq:of}-\eqref{eq:constinterval} is equivalent to an $\ell_{0}$-norm minimization.
Attaining sparse vectors by minimizing the $\ell_{0}$-norm has been shown to be NP-hard \citep{Natarajan1995}.

An MILP solver first solves the following LP relaxation:
\begin{equation}
\max\limits_{x\in [0,1],t\geq 0}\left\{\sum\limits_{i\in I_{PTV}}x_{i}:(\dot{D}t)_{i}\geq 8.5x_{i},(x,t)\in \mathcal{X}\right\}, \label{eq:lprelaxation}
\end{equation}
where $\mathcal{X}$ is a polyhedron.
Substituting $y_{i}=1-x_{i}$ yields the following equivalent program:
\begin{equation*}
\min\limits_{y\in [0,1],t\geq 0}\left\{\sum\limits_{i\in I_{PTV}}|y_{i}|:y_{i}\geq \tfrac{8.5-(\dot{D}t)_{i}}{8.5},(1-y,t)\in \mathcal{X}\right\}.
\end{equation*}
So, the LP relaxation \eqref{eq:lprelaxation} is equivalent to $\ell_{1}$-norm minimization. Literature, predominately in the field of image processing, suggests that an optimal solution for the $\ell_{1}$-norm minimization (linear relaxation of the MILP) provides a good approximation of an optimal solution for the $\ell_{0}$-norm minimization (MILP): \citet{Donoho2006} shows that, in most cases, the $\ell_{1}$-approximation is also the sparsest solution for underdetermined systems of equations. \citet{Li2010} discusses necessary and sufficient conditions for the equivalence of $\ell_{0}$-norm and $\ell_{1}$-norm solutions in general. These conditions do not hold in our case. Nevertheless, \citet{Candesetal2005} provide numerical evidence for the power of these $\ell_{1}$-approximations. So, when the MILP \eqref{eq:of}-\eqref{eq:constinterval} is solved with an MILP solver, the solver implicitly uses an approximation that is known to be accurate. This explains the results by \citet{Gorissen2013}, where an MILP solver quickly determines good treatment plans.
However, our results in Section \ref{sec:comparison} show that DOPSA still outperforms an MILP solver in finding good solutions in a short time-span, which advocates the power of DOPSA. 

Compared to MILP-based formulations, DOPSA has another advantage. When more DVH-constraints are added, the time for constraint verifications increases linearly, or sublinearly when solutions are removed in the process. Contrarily, both the number of variables and the number of constraints of the MILP formulation grow linearly in the number of DVH criteria, which slows down an MILP solver at least quadratically. For example, when the LP relaxation is solved with an interior point method, the time complexity of solving the KKT system increases at least quadratically in the number of variables. A dual solver cannot alleviate this slowdown since the number of constraints increases as well.

\section{Conclusions}
\label{sec:conclusions}
The existing dose-volume based optimization methods MILP and IPIP have not been compared before. Our results show that each method has its own advantages. IPIP is faster while MILP determines a better solution.

DOPSA, our new algorithm, is a viable alternative to both. Its main advantage is that it does not require a costly solver. The time it takes to determine a good solution is less than MILP but slightly more than IPIP. Since IPIP returns a single solution quickly but does not keep improving the solution, DOPSA eventually determines better treatment plans. This leads to two observations: first, advances in computing power will reduce the absolute time difference between both methods, making IPIP lose its edge. Second, it is possible to combine the strength of two methods and use the IPIP solution as a starting point for DOPSA. However, the additional costs for an LP solver required by IPIP and the added complexity is not justified by the slight improvement in solution time.

We have disclosed all information necessary to implement DOPSA. We hope it receives more testing by the community, and will eventually become available to treatment planners.

\section{Acknowledgments}
The authors would like to thank Elekta Brachytherapy for supplying anonymous patient data, T. Siauw (UCSF, San Francisco CA, USA) and anonymous referees for their constructive feedback.


\addcontentsline{toc}{section}{References}
\begin{thebibliography}{24}
\providecommand{\natexlab}[1]{#1}
\providecommand{\url}[1]{\texttt{#1}}
\expandafter\ifx\csname urlstyle\endcsname\relax
  \providecommand{\doi}[1]{doi: #1}\else
  \providecommand{\doi}{doi: \begingroup \urlstyle{rm}\Url}\fi

\bibitem[Baltas et~al.(2009)Baltas, Katsilieri, Kefala, Papaioannou, Karabis,
  Mavroidis, and Zamboglou]{baltas2009}
D.~Baltas, Z.~Katsilieri, V.~Kefala, S.~Papaioannou, A.~Karabis, P.~Mavroidis,
  and N.~Zamboglou.
\newblock Influence of modulation restriction in inverse optimization with
  {HIPO} of prostate implants on plan quality: Analysis using dosimetric and
  radiobiological indices.
\newblock In \emph{World Congress on Medical Physics and Biomedical
  Engineering, September 7 - 12, 2009, Munich, Germany}, IFMBE Proceedings,
  \href{http://dx.doi.org/10.1007/978-3-642-03474-9_81}{283--286}, 2009.

\bibitem[Balvert et~al.(2015)Balvert, Gorissen, den Hertog, and
  Hoffmann]{balvert2015dwell}
M.~Balvert, B.~L. Gorissen, D.~den Hertog, and A.~L. Hoffmann.
\newblock Dwell time modulation restrictions do not necessarily improve
  treatment plan quality for prostate {HDR} brachytherapy.
\newblock \emph{Physics in Medicine and Biology},
  \href{http://dx.doi.org/10.1088/0031-9155/60/2/537}{60\penalty0 (2):\penalty0
  537--548}, 2015.

\bibitem[Beli\"{e}n et~al.(2009)Beli\"{e}n, Colpaert, and {de
  Boeck}]{Belien2009}
J.~Beli\"{e}n, J.~Colpaert, and L.~{de Boeck}.
\newblock A hybrid simulated annealing linear programming approach for
  treatment planning in {HDR} brachytherapy with dose volume constraints.
\newblock In \emph{Proceedings of the 35th International Conference on
  Operational Research Applied to Health Services},
  \href{https://lirias.hubrussel.be/handle/123456789/2730}{2009}.

\bibitem[Candes et~al.(2005)Candes, Rudelson, Tao, and
  Vershynin]{Candesetal2005}
E.~Candes, M.~Rudelson, T.~Tao, and R.~Vershynin.
\newblock Error correction via linear programming.
\newblock In \emph{Proceedings of the 46th Annual IEEE Symposium on FOCS},
  \href{http://dx.doi.org/10.1109/sfcs.2005.32}{295--308}, 2005.

\bibitem[Cho et~al.(1998)Cho, Lee, Marks~II, Sutlief, and Phillips]{cho1998}
P.~S. Cho, S.~Lee, R.~J. Marks~II, S.~G. Sutlief, and M.~H. Phillips.
\newblock Optimization of intensity modulated beams with volume constraints
  using two methods: Cost function minimization and projections onto convex
  sets.
\newblock \emph{Medical Physics}, \href{http://dx.doi.org/}{25\penalty0
  (4):\penalty0 435--443}, 1998.

\bibitem[Deist(2013)]{Deist2013}
T.~M. Deist.
\newblock Fast inverse planning for {HDR} prostate brachytherapy.
\newblock Master's thesis, {Tilburg University}, Tilburg, The Netherlands,
  2013.

\bibitem[Donoho(2006)]{Donoho2006}
D.~Donoho.
\newblock For most large underdetermined systems of equations, the minimal
  l(1)-norm near-solution approximates the sparsest near-solution.
\newblock \emph{Communications on Pure and Applied Mathematics},
  \href{http://dx.doi.org/10.1002/cpa.20131}{59\penalty0 (7):\penalty0
  907--934}, 2006.

\bibitem[Ehrgott et~al.(2010)Ehrgott, \c{C}i\v{g}dem, Horst, and
  Shao]{ehrgott2010}
M.~Ehrgott, G.~\c{C}i\v{g}dem, W.~H. Horst, and L.~Shao.
\newblock Mathematical optimization in intensity modulated radiation therapy.
\newblock \emph{Annals of Operations Research},
  \href{http://dx.doi.org/10.1007/s10288-008-0083-7}{175\penalty0 (1):\penalty0
  309--365}, 2010.

\bibitem[Gorissen et~al.(2013)Gorissen, den Hertog, and Hoffmann]{Gorissen2013}
B.~L. Gorissen, D.~den Hertog, and A.~L. Hoffmann.
\newblock Mixed integer programming improves comprehensibility and plan quality
  in inverse optimization of prostate {HDR} brachytherapy.
\newblock \emph{Physics and Medicine in Biology},
  \href{http://dx.doi.org/10.1088/0031-9155/58/4/1041}{58\penalty0
  (4):\penalty0 1041--1057}, 2013.

\bibitem[Hedar and Fukushima(2006)]{HedarFukushima2006}
A.-R. Hedar and M.~Fukushima.
\newblock Derivative-free filter simulated annealing method for constrained
  continuous global optimization.
\newblock \emph{Journal of Global Optimization},
  \href{http://dx.doi.org/10.1007/s10898-005-3693-z}{35\penalty0 (4):\penalty0
  521--549}, 2006.

\bibitem[Hoskin et~al.(2007)Hoskin, Motohashi, Bownes, Bryant, and
  Ostler]{Hoskin2007}
P.~J. Hoskin, K.~Motohashi, P.~Bownes, L.~Bryant, and P.~Ostler.
\newblock High dose rate brachytherapy in combination with external beam
  radiotherapy in the radical treatment of prostate cancer: initial results of
  a randomised phase three trial.
\newblock \emph{Radiotherapy and Oncology},
  \href{http://dx.doi.org/10.1016/j.radonc.2007.04.011}{84\penalty0
  (2):\penalty0 114--120}, 2007.

\bibitem[Hoskin et~al.(2013)Hoskin, Colombo, Henry, Niehoff, Hellebust,
  Siebert, and Kov\'acs]{Hoskin2013}
P.~J. Hoskin, A.~Colombo, A.~Henry, P.~Niehoff, T.~P. Hellebust, F.-A. Siebert,
  and G.~Kov\'acs.
\newblock {GEC/ESTRO} recommendations on high dose rate afterloading
  brachytherapy for localised prostate cancer: An update.
\newblock \emph{Radiotherapy and Oncology},
  \href{http://dx.doi.org/10.1016/j.radonc.2013.05.002}{107\penalty0
  (3):\penalty0 325--332}, 2013.

\bibitem[Karabis et~al.(2005)Karabis, Giannouli, and Baltas]{Karabis2005}
A.~Karabis, S.~Giannouli, and D.~Baltas.
\newblock {HIPO}: A hybrid inverse treatment planning optimization algorithm in
  {HDR} brachytherapy.
\newblock \emph{Radiotherapy and Oncology},
  \href{http://dx.doi.org/10.1016/S0167-8140(05)81018-7}{76\penalty0
  (S2):\penalty0 S29}, 2005.

\bibitem[Kirkpatrick et~al.(1983)Kirkpatrick, Gelatt, and Vecchi]{kirkpatrick}
S.~Kirkpatrick, C.~D. Gelatt, and M.~P. Vecchi.
\newblock Optimization by simulated annealing.
\newblock \emph{Science},
  \href{http://dx.doi.org/10.1126/science.220.4598.671}{220\penalty0
  (4598):\penalty0 671--680}, 1983.

\bibitem[Lahanas and Baltas(2003)]{lahanasQ2003}
M.~Lahanas and D.~Baltas.
\newblock Are dose calculations during dose optimization in brachytherapy
  necessary?
\newblock \emph{Medical Physics},
  \href{http://dx.doi.org/10.1118/1.1580483}{30\penalty0 (9):\penalty0
  2368--2375}, 2003.

\bibitem[Lahanas et~al.(2003)Lahanas, Baltas, and Giannouli]{Lahanas2003}
M.~Lahanas, D.~Baltas, and S.~Giannouli.
\newblock Global convergence analysis of fast multiobjective gradient-based
  dose optimization algorithms for high-dose-rate brachytherapy.
\newblock \emph{Physics in Medicine and Biology},
  \href{http://dx.doi.org/10.1088/0031-9155/48/5/304}{48\penalty0 (5):\penalty0
  599--617}, 2003.

\bibitem[Lessard and Pouliot(2001)]{Lessard2001}
E.~Lessard and J.~Pouliot.
\newblock Inverse planning anatomy-based dose optimization for
  {HDR}-brachytherapy of the prostate using fast simulated annealing algorithm
  and dedicated objective function.
\newblock \emph{Medical Physics},
  \href{http://dx.doi.org/10.1118/1.1368127}{28\penalty0 (5):\penalty0
  773--779}, 2001.

\bibitem[Li(2010)]{Li2010}
Y.~Li.
\newblock Two conditions for equivalence of 0-norm solution and 1-norm solution
  in sparse representation.
\newblock \emph{Neural Networks},
  \href{http://dx.doi.org/10.1109/TNN.2010.2049370}{21\penalty0 (7):\penalty0
  1189--1196}, 2010.

\bibitem[Milickovic et~al.(2002)Milickovic, Lahanas, Papagiannopoulou,
  Zamboglou, and Baltas]{Milickovic2002}
N.~Milickovic, M.~Lahanas, M.~Papagiannopoulou, N.~Zamboglou, and D.~Baltas.
\newblock Multiobjective anatomy-based dose optimization for
  {HDR}-brachytherapy with constraint free deterministic algorithms.
\newblock \emph{Physics in Medicine and Biology},
  \href{http://dx.doi.org/10.1088/0031-9155/47/13/306}{47\penalty0
  (13):\penalty0 2263--2280}, 2002.

\bibitem[Natarajan(1995)]{Natarajan1995}
B.~K. Natarajan.
\newblock Sparse approximate solutions to linear systems.
\newblock \emph{SIAM Journal on Computing},
  \href{http://dx.doi.org/10.1137/S0097539792240406}{24\penalty0 (2):\penalty0
  227--234}, 1995.

\bibitem[Niemierko and Goitein(1990)]{niemierko1990random}
A.~Niemierko and M.~Goitein.
\newblock Random sampling for evaluating treatment plans.
\newblock \emph{Medical Physics},
  \href{http://dx.doi.org/10.1118/1.596473}{17\penalty0 (5):\penalty0
  753--762}, 1990.

\bibitem[Panchal(2008)]{Panchal2008}
A.~Panchal.
\newblock \emph{Harmony search optimization for {HDR} prostate brachytherapy}.
\newblock PhD thesis, {Rosalind Franklin University of Medicine and Science},
  North Chicago, United States, 2008.

\bibitem[Siauw et~al.(2011)Siauw, Cunha, Atamt{\"u}rk, Hsu, Pouliot, and
  Goldberg]{Siauw2011}
T.~Siauw, A.~Cunha, A.~Atamt{\"u}rk, I.~Hsu, J.~Pouliot, and K.~Goldberg.
\newblock {IPIP}: A new approach to inverse planning for {HDR} brachytherapy by
  directly optimizing dosimetric indices.
\newblock \emph{Medical Physics},
  \href{http://dx.doi.org/10.1118/1.3598437}{38\penalty0 (7):\penalty0
  4045--4051}, 2011.

\bibitem[Zaghian et~al.(2014)Zaghian, Lim, Liu, and Mohan]{zaghian2014}
M.~Zaghian, G.~Lim, W.~Liu, and R.~Mohan.
\newblock An automatic approach for satisfying dose-volume constraints in
  linear fluence map optimization for {IMPT}.
\newblock \emph{Journal of Cancer Therapy},
  \href{http://dx.doi.org/10.4236/jct.2014.52025}{5\penalty0 (2):\penalty0
  198--207}, 2014.

\end{thebibliography}
\end{document}